\documentclass[%
 reprint,
 amsmath,amssymb,
 aps,pre,floatfix
]{revtex4-2}
\usepackage{graphicx}
\usepackage{dcolumn}
\usepackage{bm}
\usepackage{xcolor}

\begin{document}
\title{Evolutionary Hysteresis: Cycling about in a Rugged Landscape}

\author{Luke Piszkin}
 \author{Dervis Can Vural}
 \email{dvural@nd.edu}
\affiliation{
Department of Physics, University of Notre Dame, Notre Dame, IN 46556, USA
}
\begin{abstract}
In this work, we integrate theoretical modeling, molecular simulation, and empirical analysis to identify and characterize evolutionary hysteresis. We first show how epistatic interactions create bistable fitness landscapes and structural hysteresis in a two-locus Wright–Fisher model, revealing two distinct hysteresis regimes under cyclic and noisy selection. Notably, an epistatically constrained population achieves maximal average fitness at an intermediate level of environmental stochasticity. We then extend this framework to more complex systems, demonstrating robust hysteresis loops in both a disordered multi-locus model and in biophysically realistic simulations of protein structural flexibility. Finally, we present direct empirical evidence of evolutionary hysteresis. By analyzing two decades of metagenomic time-series data from freshwater C. Nanopelagicaceae experiencing strong seasonal temperature cycles, we find that approximately 65\% of seasonally oscillating alleles exhibit statistically significant hysteresis. Together, these results establish hysteresis as a general, measurable feature of evolution and a potential probe of complex fitness landscapes.\\ 
\textbf{Keywords}: fitness landscape, epistasis, hysteresis, protein evolution, microbial evolution\\
\textbf{Classification:} Biological Sciences, Evolution
\\
\\
\textbf{Significance Statement:} Evolving populations are faced with adapting to changing environments, while interactions between genetic components constrain available evolutionary pathways. This leads to hysteresis, a history-dependence of evolution. We theoretically, computationally, and empirically characterize evolutionary hysteresis in simple and complex population genetic models, protein evolution with disordered epistasis, and real-world seasonal microbial evolution. Our findings demonstrate how hysteresis may serve to probe complex features of rugged fitness landscapes. 
\end{abstract}
\maketitle

\section{Introduction}
Each small step in an organism's evolution is contingent upon its history: the fitness of a novel gene is modulated by a genomic context shaped by a sum of prior environments \cite{travisano_experimental_1995,blount_historical_2008,kryazhimskiy_microbial_2014,johnson_epistasis_2023}. Genes affecting each other's fitness (epistasis) can constrain and obstruct pathways in the fitness landscape, preventing the population from keeping up with the changes of an environment in real time \cite{weinreich_perspectivesign_2005,poelwijk_reciprocal_2011,de_visser_empirical_2014}.

The history-dependent response of a physical material is known as \emph{hysteresis}. For example, when a magnetic field is swept up and down, the magnetization of the material lags behind, forming a characteristic loop \cite{jiles_theory_1984,chakrabarti_dynamic_1999}. Analogous memory effects appear in ecological, physiological, and even cultural systems \cite{osmond_evolutionary_2017,roemhild_cellular_2018,pascual_epistasis_2020}. Yet in evolutionary biology, hysteresis has not been quantified and characterized, despite its conceptual connection to historical contingency and epistasis.

In this work, we integrate theoretical modeling, numerical simulation, and empirical analysis to identify and characterize evolutionary hysteresis, and explore its biological consequences. We begin by developing an analytically tractable two-locus, two-allele model \cite{kimura_probability_1962,kimura_diffusion_1964,park_bistability_2011} that reveals how epistatic interactions generate bistable fitness landscapes and, consequently, structural hysteresis analogous to memory effects in physical systems \cite{jiles_theory_1984,chakrabarti_dynamic_1999}. We then subject this model to oscillating and noisy selection pressures to distinguish between two varieties of epistasis, dynamic and static. This analysis also uncovers a non-trivial effect that is the joint outcome of history dependence and noise: we find that an epistatically constrained population has maximal mean fitness for some finite level of environmental stochasticity, a phenomenon akin to stochastic resonance or annealing \cite{brady_optimization_1985,heo_emergence_2009,stewart_environmental_2012,schulz_epistatic_2025}.

We then extend this framework to more complex and realistic scenarios. First, by scaling up to a multi-locus multi-allele model with random pairwise epistatic couplings \cite{neidhart_adaptation_2014,boffi_how_2024,das_epistasis-mediated_2025}, we find that increasing the variance of this ``epistatic disorder'' systematically alters the area and shape of evolutionary hysteresis loops. This result demonstrates how hysteresis is a feature of both the strength and disorder of epistatic couplings. 

Next, we study evolutionary hysteresis in a molecular context, where we perform biophysically realistic \textit{in silico} evolution experiments. We cyclically select protein sequences for high and low structural flexibility, and plot the realized physical flexibility throughout the cycle. Our simulations confirm that local epistatic constraints among amino acid residues \cite{starr_epistasis_2016,domingo_causes_2019,husain_physical_2020} produce clear hysteresis loops in the protein's adaptive trajectory, bridging the gap between abstract models and molecular evolution.

Then, we demonstrate how key evolutionary parameters such as epistatic strength, environmental cycling rate, and selection noise jointly govern the morphology of these hysteresis loops. As such, our work paves the way for using hysteresis as a novel experimental probe to infer the features of an underlying fitness landscape. 

Lastly, we test the predictions of this framework empirically. We analyze two decades of metagenomic time-series data from a freshwater bacterium, \textit{Candidatus Nanopelagicaceae}, inhabiting a lake with strong seasonal temperature cycles \cite{rohwer_two_2025}. By tracking thousands of seasonally oscillating alleles, we find that a remarkable 65\% display statistically significant hysteresis loops when their frequency is plotted against water temperature. This analysis provides, to our knowledge, the first direct empirical evidence of evolutionary hysteresis.

From population genetics models to biophysically realistic proteins and empirical field data, our integrated approach establishes hysteresis as a general and measurable feature of evolution in cyclic and fluctuating environments.

\section{Results}
 To introduce epistasis into a two-locus two-allele Wright-Fisher model (e.g. $a_1b_1, a_1b_2, a_2b_1, a_2b_2$), we define a simple fitness landscape
\begin{equation}
    \omega_{11} = 1+2s, \hspace{2mm} \omega_{12} = \omega_{21} = 1+s-J,\hspace{2mm}  \omega_{22} = 1
    \label{fitness_landscape}
\end{equation}
where we assign a relative fitness benefit $s$ to genotypes containing alleles of type-1, with an epistatic correction $J$. Such constructions are standard in population genetics and allow for precise control of evolutionary parameters \cite{karlin_general_1975}. The case $s < J$ is known as reciprocal sign epistasis, where the ``mixed'' genotypes ($a_1,b_2$) or $(a_2,b_1)$ is less fit than the ``pure'' genotypes ($(a_1,b_1)$ and $(a_2,b_2)$). Reciprocal sign epistasis may be caused by functional and physical incompatibilities of biomolecules, such as charge mismatches of amino acids that disrupt protein folding or enzymatic sites \cite{starr_epistasis_2016, domingo_causes_2019,kvitek_reciprocal_2011}.

A standard diffusion approximation (Supporting Information, SI. I) yields a Fokker-Planck equation governing the probability density $P(p_a,p_b)$ of frequencies $p_a,p_b$ of genotypes $a_1,b_1$ (that of others are $p_a'=1-p_a$ and $p_b'=1-p_b$),
\begin{align}
   \partial_t P = -\sum_i\partial_{p_i}(A_iP) + 
   \frac{1}{2}\sum_{i,j}\partial_{p_i}\partial_{p_j} (B_{ij}P)
   \label{2D_fokker_probability}
\end{align}
where $A(p_a,p_b)$ and $B(p_a,p_b)$, are the advection and diffusion coefficients, and depend explicitly on the model parameters $s$, $J$, mutation rate $u$, and population size $N$. From Eqn.\ref{fitness_landscape} and Eqn.\ref{2D_fokker_probability}, one obtains a potential solution of the stationary state of the population distribution for a parameter set $(s,J,u,N)$ of the form
$P_{st}(\vec{p})\propto e^{-2N\psi(\vec{p})}$, where the $\psi(\vec{p})$ is called the fitness potential and is given by
\begin{align}
    &\psi = (2u-1)\log[\bar{\omega}(p_a,p_b)] -  
     2u\log[p_ap_b(1-p_a)(1-p_b)]\nonumber\\
     &\bar{\omega} = \omega_{11}p_ap_b + \omega_{12}p_ap_b' + 
    \omega_{21}p_bp_a' + \omega_{22}p_a'p_b'\label{2D_wbar}
\end{align}
where $\bar{\omega}$ is the average fitness. The structure of the fitness potential is governed by the evolutionary parameters $s,J,u$. 
\begin{figure*}[hbtp]
    \centering
    \includegraphics[width=1.0\linewidth]{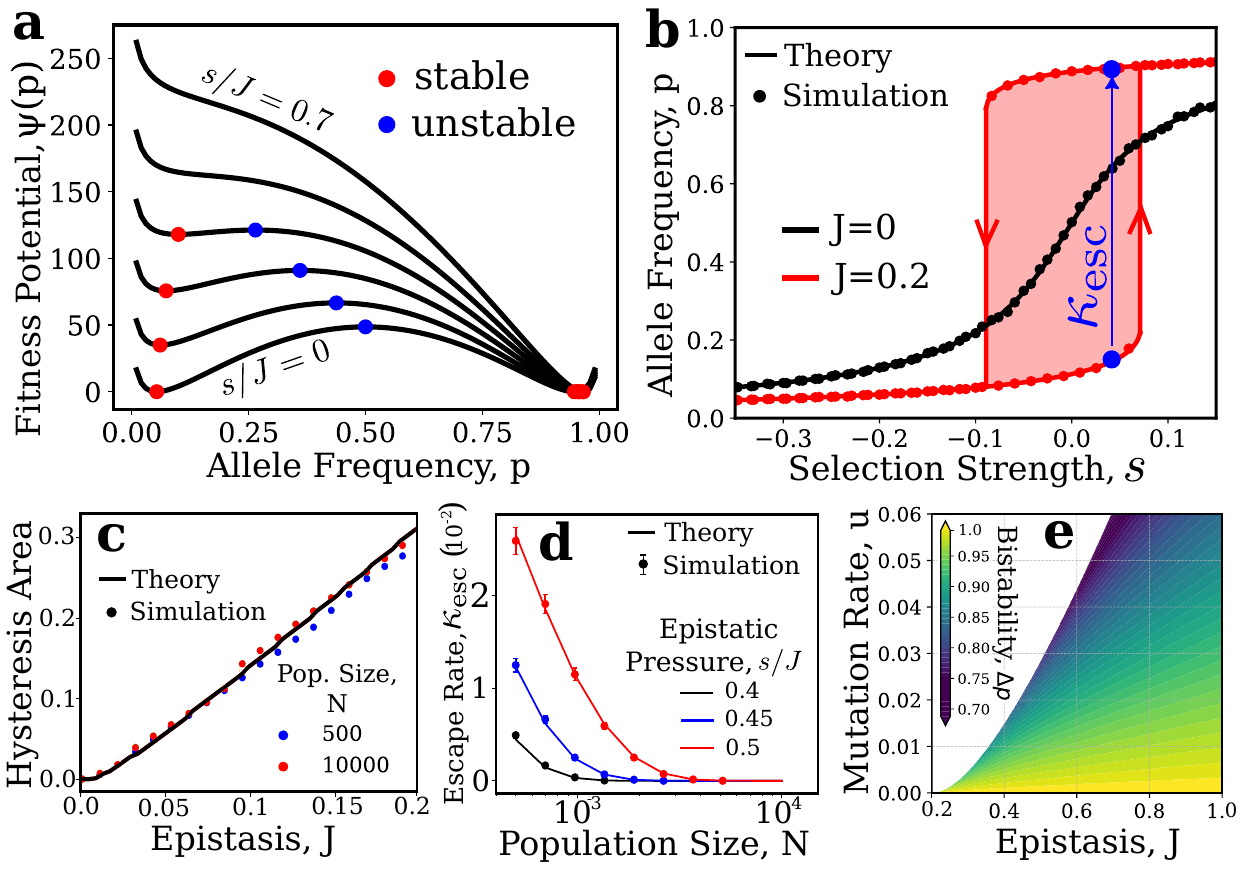}
    \caption{\textbf{Hysteresis in the two-locus two-allele Wright-Fisher model with epistasis}. (a) The effective fitness potential of an epistatic population (Eqn.\ref{1D_eff_potential}) turns from a single to a double-well form when there is significant reciprocal sign epistasis $(s<J)$ for mixed genotypes. (b) In our simulations, allele frequencies exhibit a hysteresis loop as we sweep selection strength back and forth ($J=0.2$, red circles). The hysteresis loop area vanishes in the absence of epistasis ($J=0$, black circles). In (b), (c) The form of the curves and hysteresis area matches our analytical formulas (Eqn.S2) for populations as small as $N=500$ (solid curves). (d) The demographic noise in smaller populations allow random jumps from one hysteresis branch to the other (i.e. between two potential minima). We determine this escape rate $\kappa_{\textrm{esc}}$ from the lower to upper stable allele state via simulations (circles) and analytical theory (Eqn.\ref{kramers_rate}, solid curves). (e) The phase diagram measuring the distance between the potential wells (blue to yellow). The white region is where the potential becomes single welled and hysteresis vanishes.}
    \label{fig:1}
\end{figure*}
Due to the symmetric nature of Eqn.\ref{fitness_landscape} the stable points of $\psi$ fall along the line $p=p_a=p_b$, and the system can be reduced to one dimension, Eqn.\ref{1D_eff_potential}.
\begin{equation}
    \psi(p) = -(1-2u)\ln[\bar{\omega}(p)] - 2u\ln[p(1-p)]
    \label{1D_eff_potential}
\end{equation}
This fitness potential shows us how varying $s$ can cause a transition between a mono-stable and bistable system \cite{park_bistability_2011}(Fig.\ref{fig:1}a,b). 

Because of the epistatic fitness barrier, reversing the sign of $s$ will not immediately erase the previous state, thus an evolving system will ``remember'' its local minimum and retain this state even if it no longer optimally fit. The red curve in Fig.\ref{fig:1}b is the characteristic hysteresis loop that is common to many history dependent systems. Fig.\ref{fig:1}e shows the degree of bistability (as quantified by the distance between wells) as a function of epistasis strength and mutation rate. Hysteresis does not occur in the white zone.

To examine the hysteresis behavior of epistatic populations, we simulated the evolution of a population of $N$ governed by a Wright-Fisher process with mutation and epistatic selection defined by Eqn.\ref{fitness_landscape}. The cycling of the selection pressure $s$ resembles many common scenarios in nature, such as seasonal oscillations of light, temperature, nutrient and water availability \cite{yu_distinct_2022}.

From these results, we can evaluate the quality of the diffusion approximation as a function of $N$ and $u$. The diffusion approximation accurately describes the roughly linear relationship between epistatic strength and hysteresis area (Fig.\ref{fig:1}c, Eqn.S2), and holds for populations as small as $N = 500$. 

The setup Eqn.\ref{2D_fokker_probability} assumes that the alleles at each locus are sampled independently after selection and mutation, thus are unlinked. However, we also determine the hysteresis area when the linkage equilibrium is broken (SI.II). We find that while hysteresis area is zero for fully-linked alleles, it rapidly increases and saturates to its linkage equilibrated value, with increased recombination rate (Fig.S2). 

\section*{Escape Rate over an Epistatic Fitness Barrier}
We next obtain the rate at which a population escapes over an epistatic fitness barrier. This rate measures the population’s \emph{adaptive mobility} or \emph{evolvability}: the likelihood of breaking free from an epistatic constraint to discover a new, fitter region of the landscape. Stronger epistasis or larger populations suppress this rate, rendering adaptation sluggish even when environmental conditions favor transition \cite{weissman_rate_2009}. We use Eyring-Kramer theory to obtain the barrier escape rate $\kappa_{esc} \propto \sqrt{U(x)''(x_0)|U''(x_s)|}\exp [U(x_0) - U(x_s)]$, where $x_0$ and $x_s$ are the location of the minima and barrier of a generalized potential function $U(x)$ \cite{eyring_activated_1935,kramers_brownian_1940}. Recent work has derived the correct expression for the Kramers' rate for cases with state-dependent noise \cite{rosas_kramers_2016}, which we use to calculate the escape rate over an epistatic barrier. The scaling of the escape rate with $s,J,N,$ and $u$ is given by Eqn.\ref{kramers_rate}, with a negligible prefactor (SI. III).
\begin{align}
    \kappa_{esc} \propto N\left[1 -(s-J)^2/(2J)\right]^{N(1-2u)}
    \label{kramers_rate}
\end{align}
Fig.\ref{fig:1}d shows how this rate decreases exponentially with increasing population size. For large populations, barrier crossing events become exceedingly rare, in agreement with previous results on stochastic tunneling \cite{weissman_rate_2009,altland_rare_2011}, but now with explicit dependence on the selection and epistasis parameters $s$ and $J$. As the selection strength increases, escape events become much more common. These results suggest that environmental conditions and demographic constraints jointly determine whether populations remain trapped behind epistatic barriers or traverse them.

\section*{Stochastic Environments Anneal History Dependence}
Having established how static epistatic barriers shape adaptation, we next ask how fluctuating environments and stochastic selection modulate these effects. Populations often experience rapidly changing environments where selection fluctuates faster than adaptation can respond \cite{mustonen_fitness_2009,nguyen_environmental_2020}. 
In such evolutionary “seascapes,” the joint effects of epistasis and environmental stochasticity remain poorly understood \cite{bank_epistasis_2022}. 
Environmental noise can either hinder or facilitate adaptation by altering the balance between robustness and evolvability \cite{steiner_environmental_2012,melbinger_impact_2015}. 
For instance, moderate fluctuations may help populations escape local fitness traps imposed by epistasis, whereas with excessive variability populations cannot keep up, and will experience a fitness shortfall \cite{brady_optimization_1985,heo_emergence_2009}. Understanding how noise interacts with epistatic constraints is therefore essential for predicting evolutionary outcomes in realistic, time-varying environments. 

To do so, we simulated the evolutionary trajectories of our epistatic two-locus two-allele population subject to a selective force with a cyclic component with amplitude $A$ and frequency $\Omega$ (to yield hysteresis) and a random Gaussian component $\eta(t)$ (that will modulate the hysteresis),
\begin{equation}
    s(t) = A\sin(\Omega t) + \eta(t)
    \label{oscillating_s}
\end{equation}

\begin{figure}[hbtp]
    \centering
    \includegraphics[width=1.0\linewidth]{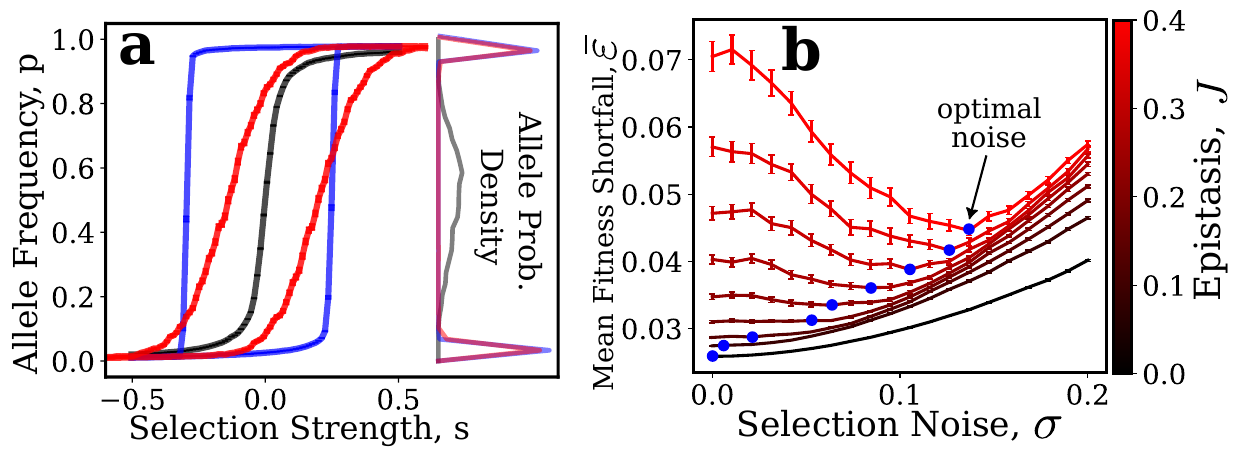}
    \caption{\textbf{Noise Anneals Hysteresis}. (a) The evolutionary response of populations in oscillating and noisy selection, $s(t) = A\sin(\Omega t) + \eta(t)$ with $A = 0.5$, $\omega = 0.001$ for (red) $J = 0.4$ with optimal noise, (blue) $J = 0.4$ with no noise ($\sigma = 0)$ and (black) $J = 0$ with no noise. As we see, the addition of noise (red) can pull a hysteresis loop (blue) closer toward a non-epistatic trajectory (black). Noise enables jumps over the epistatic barrier, thereby increasing the evolvability of the population. Panel (b) quantifies the improvement in fitness due to this change in evolvability. The blue points in indicate optimal noise that maximizes fitness in oscillatory selection conditions. Naturally, the population needs no help from noise in the absence of epistasis (black curve, $J=0$).}
    \label{fig:2}
\end{figure}
We observe that noise anneals hysteresis by enabling random jumps over epistatic barriers: Epistatic hysteresis curves (\ref{fig:2}a, blue), in the presence of moderate noise (red), approach non-epistatic ones (black), indicating enhanced evolvability (Fig.\ref{fig:2}a). However, as the noise is increased further, the population is unable to keep up with an environment fluctuating too wildly, and will suffer a fitness shortfall. Fig.\ref{fig:2}b shows the sweet spot (blue markers) between evolvability and unpredictability. The fitness shortfall $\bar{\varepsilon}$ is defined as the deviation from the ideal fitness $\bar{\omega}^*(s(t))$ (the fitness, had there been no epistatic obstruction, i.e. with $J=0$ substituted in Eqn.\ref{2D_wbar}).
\begin{equation}
    \bar\varepsilon = (1/T)\int_0^T |\bar{\omega}^*(s(t)) - \bar{\omega}(t)|dt.
    \label{fitness_shortfall}
\end{equation}
where $\bar{\omega}(t)$ is the instantaneous mean fitness of the population at time $t$ (Eqn.\ref{2D_wbar}), and $T$ is the averaging time over many cycles (cf. Methods).

As expected, the optimal value of noise is a function of the ruggedness of the fitness landscape, which in our model is quantified by $J$ (Fig.\ref{fig:2}). The higher the epistatic barrier, the larger the noise the population will benefit from.

To estimate the value and ``specificity'' of the optimal noise, we fit a quadratic model $\bar{\varepsilon}(\sigma) = \kappa\sigma^2 + b\sigma + c$ by non-linear least-squares minimization to estimate the optimal noise value $\sigma|\min(\bar{\varepsilon}) = -b/2\kappa$, along with the curvature $\kappa$: a larger specificity $\kappa$, indicates a tighter range of optimal noise. We find that specificity $\kappa$ increases with $J$ (Fig.S3), i.e., populations with stronger epistasis are less tolerant to deviations from their optimal noise amplitude. This behavior arises because larger $J$ creates steeper effective fitness wells around the adaptive optimum.

These fit parameters quantify the effect of epistasis $J$ on the mean fitness shortfall in stochastic environments (Fig.S3). For small $J$, environmental noise leads to a larger fitness shortfall as expected, since the population cannot adapt quickly to a particular selection condition before it changes. The optimal noise $\sigma|\min(\bar{\varepsilon})$ increases roughly linearly with $J$.

\section*{Dynamical Hysteresis}

If a single potential well moves between multiple points at a rate faster than the system can respond, the system will effectively experience multiple wells. This phenomenon is known as dynamical hysteresis \cite{jung_scaling_1990}. Here we quantify dynamical hysteresis in an evolutionary system, and determine how the rate of cycling $\Omega$ affects hysteresis features. 

\begin{figure}[hbtp]
    \centering
    \includegraphics[width=1.0\linewidth]{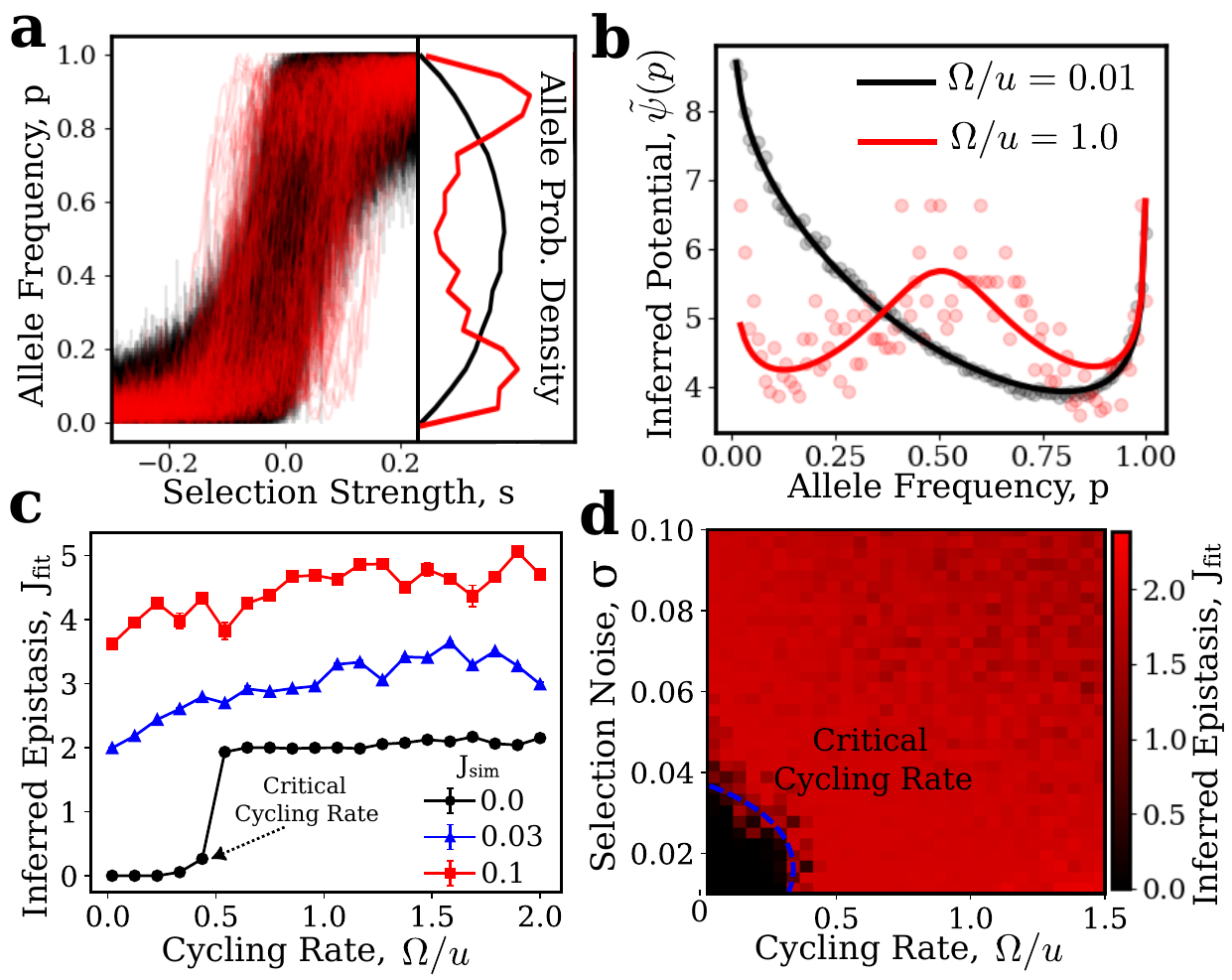}
    \caption{\textbf{Exploring dynamical hysteresis.} A single-well potential well moving back and forth quick enough will have the effect of a double well, leading to similar hysteresis effects, remarkably, even in the absence of epistasis ($J=0$). (a) As we increase oscillation frequency from $\Omega / u = 0.01$ (black) to $\Omega/u = 1.0$ (red), we observe the emergence of dynamic hysteresis, and the bistability that underlies it. (b) We use the probability density of alleles (inset of (a)) to fit an effective fitness potential via Eqn.\ref{1D_eff_potential}. As we increase the cycling frequency, an effective double-well structure emerges, mimicking the role of epistasis. (c) Regardless of the true value of epistasis (black: $J_\text{sim}=0$, blue: $J_\text{sim}=0.1$, red: $J_\text{sim}=0.2$), we can fit an ``inferred effective epistasis'' $J_\text{fit}$ to the emergent effective potential. For $J=0$ we observe a phase transition where hysteresis emerges once the cycling rate exceeds the evolutionary relaxation rate $(\omega / u)_\textrm{crit}$. This transition softens and gradually vanishes with larger values of true epistasis $J_\text{sim}$. (d) A phase diagram for $J=0$ shows the interplay between selection noise $\sigma$ and cycling rate, and delineates the regime where dynamical hysteresis emerges.} 
    \label{fig:3}
\end{figure}

We indeed observe the emergence of dynamical hysteresis in the fast cycling regime, even in the absence of epistasis, $J=0$ (Fig.\ref{fig:3}a). To extract an effective double-well fitness potential, we fit our simulation results of allele frequency distributions to our theoretical $P_{st}(p)=e^{-2N\psi(p)}$, with Eqn.\ref{1D_eff_potential} and find good agreement (Fig.\ref{fig:3}b). This fit allows us to infer and effective epistatic parameter $J_\textrm{fit}$ that indicates the onset of dynamical hysteresis for evolving populations. 

In the absence of true epistasis ($J=0$) we observe a phase transition: for low cycling rates there is no hysteresis (and $J_\textrm{fit}=0$), but as we hit a critical cycling rate, hysteresis emerges (and $J_\textrm{fit}>0$) (Fig.\ref{fig:3}c). As true epistasis $J$ is increased, the transition point softens, and then vanishes. 

Next, we investigate the impact of selection noise $\sigma$ on dynamical hysteresis, for $J=0$. As selection noise increases, the critical cycling rate decreases, leading to a larger $J_{\text{fit}}$ at lower $\Omega$ (Fig.\ref{fig:3}d). We observe a regime where small noise and fast cycling leads to greater inferred epistasis, which then lessens as the noise is increased. 

\section*{Hysteresis in Protein Evolution}
The two-locus Wright-Fisher model revealed how hysteresis emerges as a consequence of epistasis. However, it is not obvious whether hysteresis is a sensitive phenomena specific to this simple model, or it is a robust, generic property exhibited by realistic genomes where thousands of genes couple together through a complex network of epistatic interactions. For example, should we expect a large number of random (positive and negative) interactions $J_{ij}$ between gene pairs $i,j$ to ``cancel out'', and wash away hysteresis? Moreover, will the physical, structural properties of a protein, which might not be simply a function of pairwise interactions, exhibit similar hysteresis loops?

\begin{figure}[htbp]
    \centering
    \includegraphics[width=1.0\linewidth]{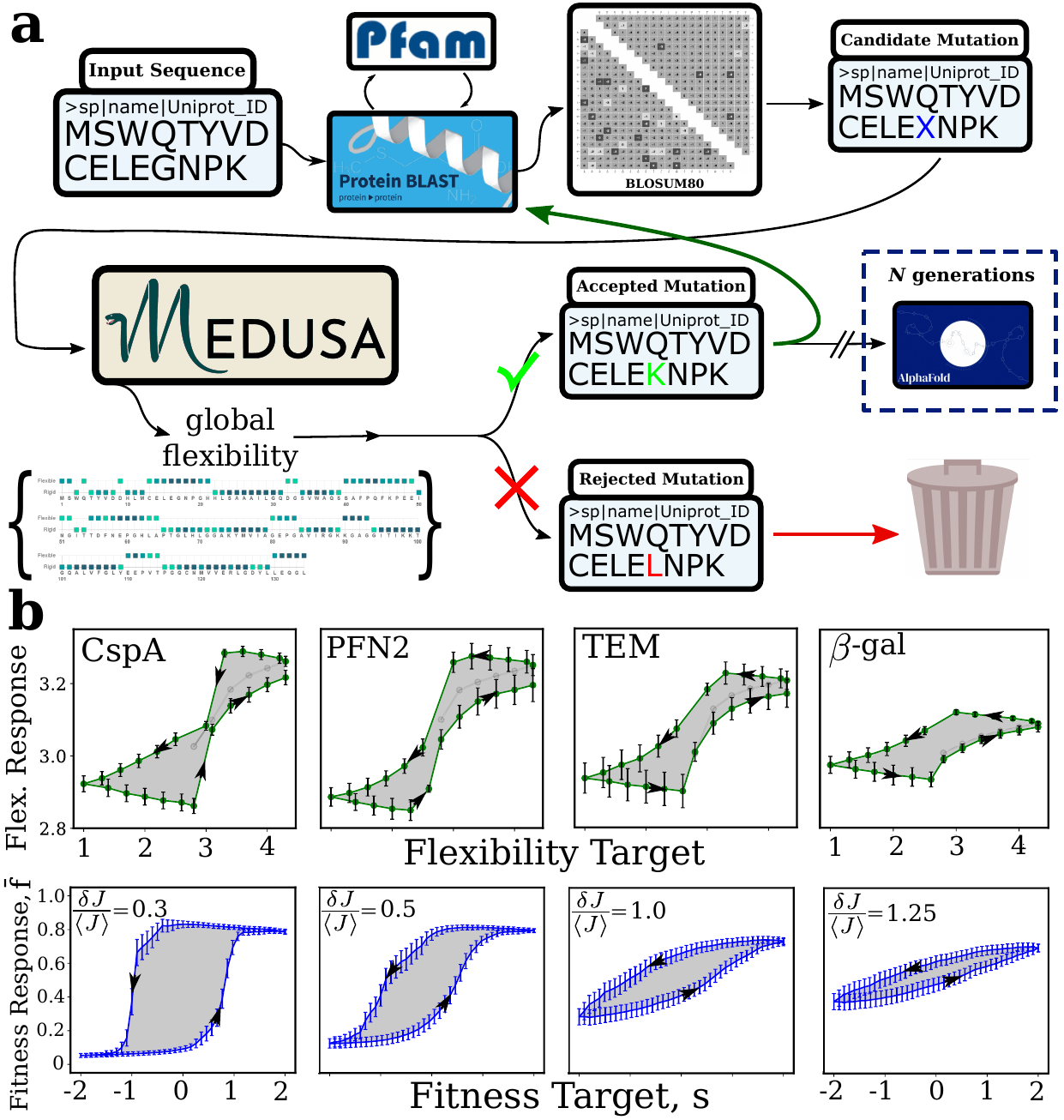}
    \caption{\textbf{Evolutionary hysteresis in biophysically realistic proteins and disordered epistatic networks}. (a) Method for evaluating hysteresis in protein flexibility evolution using the Protein Evolution Parameter Calculator (PEPC). (b) \textbf{Top row:} PEPC simulations performed on the sequences of Cold Shock Protein A (CspA), Profilin-2 (PFN2), TEM $\beta$-lactamase (TEM), and $\beta$-galactosidase ($\beta$-gal) display hysteresis loops of varying sizes and morphologies. Mechanical interactions between amino acids govern the physical flexibility of the protein, which we select for. These interactions obstruct evolutionary paths during selective cycles. \textbf{Bottom row:} Hysteresis loops for a $L$-locus $K$-allele Wright-Fisher system where the alleles interact according to a disordered epistasis network $J_{ij}$. Simulations with $L=50,K=20$ and varying epistatic disorder $\delta J / \langle J\rangle$, evolving under quasi-static cycling of selection conditions $s$ are shown. While an average value of zero $\langle J\rangle=0$ yields finite loop area (Fig.S5) increasing the disorder $\delta J$ constricts the loop.}
    \label{fig:4}
\end{figure}

We answer both questions by demonstrating the existence of hysteresis loops, first, in a complex disordered L-locus K-allele model, and then, by running biophysically realistic molecular evolution simulations, where residue–residue interactions arise from the biophysics of protein folding and structural flexibility (Fig.\ref{fig:4}).

For the L-locus K-allele model, the epistatic interaction matrix $J_{ij}^{\alpha,\beta}$ is populated from a Gaussian with mean $\langle J \rangle$ and variance $\delta J$, corresponding to increased ruggedness of the fitness landscape. The fitness potential for this more complex model is obtained through the same procedure outlined above (SI. IV),
 \begin{align}
    \psi_{lj}(\vec{p}) &=  \left(u\frac{K}{K-1}-1\right)\ln[\bar{\omega}]
    - \frac{u}{K-1}\ln [p_{li}(1-p_{li})^{K-1}] \nonumber\\
    \bar{\omega}(\vec{p})&= L - 1 + \sum_{\alpha, i} p_{\alpha i}s_i^\alpha -\sum_{i,j, \alpha\neq\beta} p_{\alpha i}p_{\beta j}J_{ij}^{\alpha,\beta}.\label{KL}
\end{align}
where $\vec{p} = (p_{11},p_{12},...,p_{LK})$, and the indices $\alpha,\beta$ and $i,j$ enumerate loci and alleles. We have observed clear hysteresis loops for $\langle J \rangle$ as small as 0, and for disorder as high as $\delta J/\langle J \rangle \sim 100\%$ (Fig.\ref{fig:4}b, Fig.S5). 

In our biophysically-realistic \emph{in silico} molecular evolution experiments, we considered four distinct proteins representing diverse folds and functions (Cold Shock Protein A, Profilin-2, $\beta$-lactamase, and $\beta$-galactosidase). Along the evolutionary trajectory, we mutated the proteins according to an empirically measured residue transition matrix, verified the viability of the fold (Fig.S4), and determined the physical flexibility of the folded structure (Fig.\ref{fig:4}a, Methods). Each protein was subject to cycles of selection for high and low structural flexibility. All four proteins exhibited clear hysteresis loops, confirming that local, biophysical interactions robustly produce global evolutionary memory (Fig.\ref{fig:4}).

The disordered L-locus K-allele model and molecular evolution experiments illustrate that evolutionary hysteresis is not a fine-tuned phenomena sensitive to system parameters. It is a general, reproducible feature that emerges in both simple and complex fitness landscapes.

\section*{Empirical Evidence to Evolutionary Hysteresis}
\begin{figure}[htbp]
    \centering
    \includegraphics[width=1.0\linewidth]{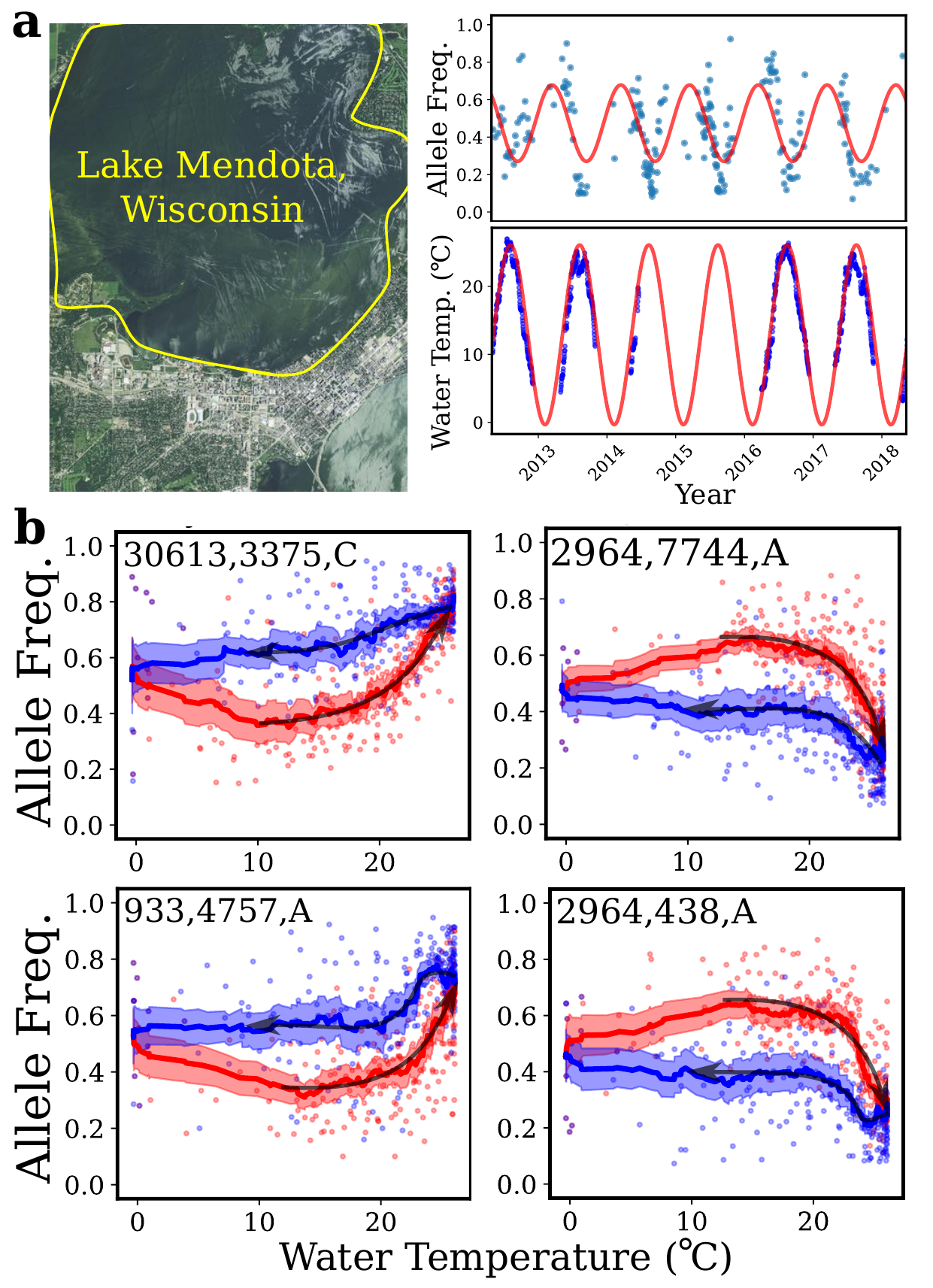}
    \caption{\textbf{Empirical evidence of hysteresis in seasonal evolution of microbial populations.} (a) We identified seasonally oscillating in alleles bacterial populations from Lake Mendota, Wisconsin \cite{rohwer_two_2025}. Top: abundance of one such allele over time from mid 2012 to 2018. Bottom: Water temperature of Lake Mendota over the same time period fit to a sinusoidal model. For both panels, dots: data, lines: sinusoidal fit with a period of one year. (b) Seasonal hysteresis loops of four seasonally oscillating alleles. In each plot, the red points indicate allele frequency in samples taken before July 31st of each year when the lake reaches is annual maximum water temperature, and the blue points indicate samples taken after July 31st of each year, representing the warming and cooling stages of seasonal evolution for the bacterial populations. The solid lines are moving averages with shaded 95\% confidence interval bands.}
    \label{fig:5}
\end{figure}

To see whether evolutionary hysteresis occurs in the wild, we analyzed genomic data from a two-decade study of bacterial populations in Lake Mendota, Wisconsin (Fig.\ref{fig:5}a) \cite{rohwer_two_2025}. We obtained a representative metagenomically-assembled genome (MAG) of the \textit{Candidatus Nanopelagicaceae} species from the NCBI Bioproject PRJNA1158976 (Assembly GCA-044348695.1), along with the filtered metagenomic reads for each sampled date from the JGI Data Portal. We mapped the reads to the representative MAG using BBMap (version 39.33, \cite{bushnell_bbmap_2014}). With the reads mapped, we used the metagenomic analysis program inStrain  (version 1.10.0, \cite{olm_instrain_2021}) to identify single nucleotide variants (SNV) with a minimum read ANI of 93\% in the species for each sample date from the years 2000 to 2019. We filtered the resulting SNVs to only biallelic entries (allele\_count = 2) that were present in more than 80\% of the sample dates. We used a Lomb-Scargle method from the astropy Python package to identify alleles with seasonally oscillating frequencies by filtering the periodgrams by those with a peak stronger than a 1\% false-positive threshold and within a oscillation period of 365 $\pm$ 30 days. This resulted in 16637 seasonal alleles of \textit{C. Nanopelagicaceae}. 

To determine whether the seasonal allele displayed hysteresis, we used the lake water temperature as a proxy for seasonal selection. We obtained the daily water temperature data for Lake Mendota from the Environmental Data Initiative \cite{magnuson_north_2024} and fit the data to a sinusoidal model (Fig.\ref{fig:5}a). For each year, we separated allele frequency sample points into two groups: before the maximum annual temperature (July 31st) and after, to obtain separate curves for the spring ``warming'' and fall ``cooling'' evolutionary periods. From these points, we calculated the 28 day moving average and 95\% confidence bands of allele frequencies and plotted them as a function of water temperature. Finally, we calculated the hysteresis area of the allele frequencies versus water temperature moving averages by the trapezoid method and estimated uncertainties from 1000 non-parametric bootstrap simulations. We determined that an allele had a non-zero hysteresis area by whether it's lower 95\% CI uncertainty was larger than zero.

From this analysis, we found that 65\% of seasonal alleles displayed significant hysteresis loop area in response to water temperature (Fig.\ref{fig:5}). These empirical hysteresis loops closely mirror those predicted by our models.

\section*{Hysteresis as an Experimental Probe}
\begin{figure}[htbp]
    \centering
    \includegraphics[width=1.0\linewidth]{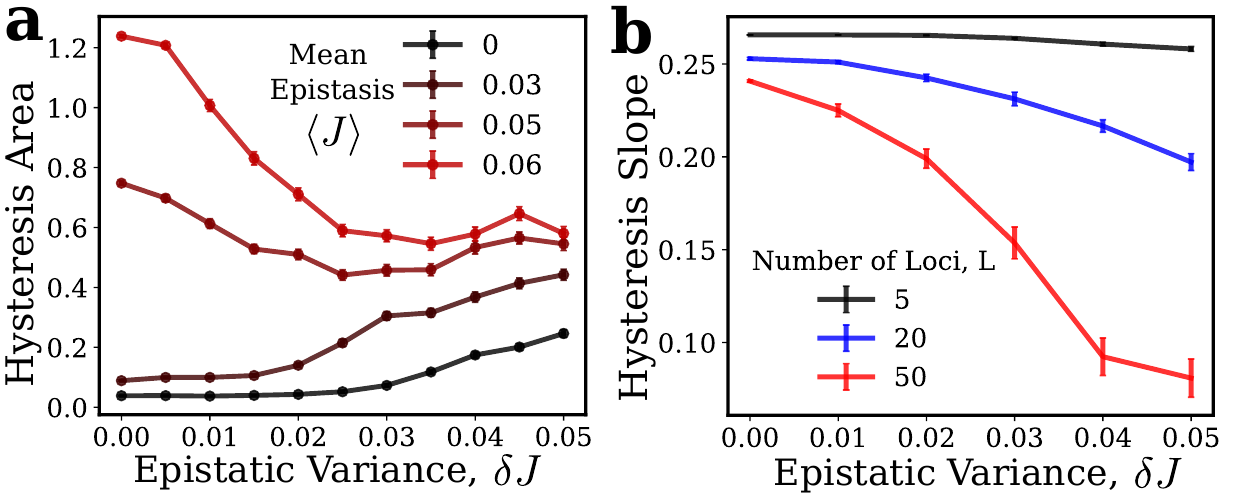}
    \caption{\textbf{Hysteresis Loop Morphology} (a) We analyzed the area of hysteresis loops for the L-locus K-allele model versus increasing epistatic variance $\delta J$ for several curves of mean epistasis $\langle J \rangle$, with $L=20$, $K=20$, $u=0.01$, and $N=1000$. At lower values of mean epistasis $\langle J \rangle$, we see an increase of hysteresis loops area with increasing epistatic variance $\delta J$. This trend is reversed for larger $\langle J \rangle$, where increasing $\delta J$ decreases the hysteresis area. (b) The average slope of hysteresis loops for increasing epistatic variance $\delta J$ for varying number of loci $L$, with $K=20$, $\langle J \rangle = 0.02$, $u = 0.002$, and $N=1000$. The slope of hysteresis loops decreases with increasing $\delta J$, and this effect is magnified by increasing the number of loci $L$. The error bars for (a) and (b) represent the standard deviation of 96 replicate simulations. These trends anticipate how one might extract meaningful information about the epistatic landscape from hysteresis loops in complex systems.}
    \label{fig:6}
\end{figure}

Having established that hysteresis emerges generically in both minimal and disordered epistatic systems, and out in the wild, we next ask whether the shape and size of the loop can serve as a probe to measure the evolutionary parameters of a real system. We show that by performing cyclic selection empirically, one can gain insights into $K$, $L$, $\langle J\rangle$, $\delta J$ in real systems (Fig.\ref{fig:6}).

To map the loop morphology to underlying evolutionary parameters, we expand the effective potential $\psi$ (Eqn.\ref{KL}) around the symmetric state $p_{li}=1/K$ in terms of the mean allelic deviation $m=\frac{1}{L}\sum_{l,i}(p_{li}-1/K)$, obtaining a coarse-grained potential $\phi(m)=a_2 m^2+a_4 m^4-hm$, with coefficients $a_2\simeq \frac{u}{K-1}\chi_K-\frac{1-u/K}{K-1}\lambda_2(J,L)+\alpha\,\delta J^2$ and $h\simeq \frac{1-u/K}{K-1}\lambda_1(L)s$, where $\chi_K$ arises from the entropic curvature of the mutation term and $\lambda_{1,2}$ describe the additive and pairwise contributions to $\ln\bar{\omega}(p)$. Hysteresis occurs when $a_2<0$, i.e., when the epistatic stabilization $\lambda_2(J,L)$ dominates mutation and disorder, and the barrier height between the wells approximately scales as $\Delta\phi\sim a_2^2/(4a_4)$. 

Because $a_2$ and $h$ both carry the prefactor $(1-u/K)/(K-1)$, increasing the allelic diversity $K$ effectively rescales both the curvature and field terms, reducing the curvature of $\psi$ near the symmetric state. In the large-$K$ limit, the entropic mutation term $-(u/(K-1))\ln p_{ij}$ of the fitness potential becomes negligible, flattening $\psi$ near the $p_{ij}\ll1$ region.  
This entropic term decreases approximately as $1/(K-1)$, corresponding to only a small change relative to typical epistatic couplings ($\delta J\sim10^{-2}$).  
Thus, increasing $K$ produces slightly broader and flatter loops, consistent with simulation results (Fig.S5).

Varying the mean coupling $\langle J \rangle$ directly modifies the stabilizing curvature $a_2\!\propto\!-\langle J \rangle(L-1)/(K-1)$.  
Larger $\langle J \rangle$ therefore deepens the bistable wells and synchronizes switching across loci, yielding wider and more squared loops (Fig.S5).  
By contrast, increasing the epistatic disorder $\delta J$ adds a positive correction to $a_2$, effectively reducing bistability.  
At small $\langle J \rangle$, disorder can enhance loop area by introducing additional metastable states, whereas at large $\langle J \rangle$ it blurs the transition and decreases loop area and slope, as observed in simulations (Fig.\ref{fig:6}a).

For the number of loci $L$, the mean epistatic contribution to $\bar{\omega}$ grows approximately as $\langle J \rangle(L-1)$, but the local fields acting on each locus,
$h_{\alpha i} = s_i^\alpha 
- \sum_{\beta\neq\alpha}\sum_j p_{\beta j}\,J_{ij}^{\alpha,\beta}$,
acquire a disorder width
$\mathrm{SD}[h] \simeq \delta J \sqrt{(L-1)\,\bar{H}}, 
\hspace{2mm} \bar{H} = \sum_i p_{\beta i}^2$.
As $L$ increases, the variance of the effective fields broadens as $\sqrt{L}$, while the mean epistasis grows linearly with $L$. This causes loci to switch at increasingly diverse values of the driving field, producing more gradual hysteresis loops with increasing $L$ (Fig.\ref{fig:6}b, Fig.S5).

Overall, these scaling relations rationalize the observed simulation trends (Fig.\ref{fig:6}) and suggest that observable features of evolutionary hysteresis loops provide information about the structure of the epistatic landscape.

In addition to loop area and morphology measurements, dynamical hysteresis can also probe the epistasis: A cyclic selection experiment can be performed to detect epistasis by measuring $J_\text{fit}(\Omega)$ and quantifying the sharpness of the phase transition.

In principle, the full shape of the loop, including branch curvature, asymmetry, and minor-loop closure, should contain enough information to reconstruct not only coarse descriptors such as barrier height and disorder, but also finer details of the fitness landscape and the epistatic network. Solving this inverse problem would require incorporating the complete loop trajectory rather than summary quantities such as area or squareness, as well as accounting for mutation rates, linkage, and recombination that may complicate the procedure. A promising direction could be developing analytic or machine-learning surrogates that map loop morphologies to landscape statistics. This approach could turn hysteresis measurements into a new quantitative probe of epistasis networks and fitness landscapes, but its realization lies beyond the scope of the present work.

\section{Discussion}
We have theoretically and empirically demonstrated evolutionary hysteresis, explored its dependence on epistasis, mutation rate, and environmental fluctuations, characterized its structural and dynamical varieties, and found evolutionary hysteresis in seasonally varying alleles in real bacterial populations. These efforts represent an important step toward our ability to model and evaluate complex evolutionary trajectories in real populations. 

Critical evaluation of how hysteresis relates to epistatic strength, environmental noise, cycling rate, and epistatic disorder may provide valuable insights for further characterization of the complex genotype-phenotype-fitness maps for many systems. We will be interested in seeing if future experimental studies can implement selective cycles, measure hysteresis loops, and use the shape and area of the loop to probe epistatic interactions, or to chart fitness landscape features \cite{milocco_is_2020,moulana_genotypephenotype_2023,nguyen_ba_barcoded_2022,chen_sampling_2024,dichio_statistical_2023}.

We have found the area and shape of hysteresis curve to depend on epistatic strength, environmental noise, cycling rate, and epistatic disorder in simple models; we have also seen, in more realistic protein evolution simulations, that the shapes of these loops can have complex, irregular shapes. Because these simulations operate on explicit molecular structures and not abstract fitness functions, they bridge population genetics and biophysics, establishing a direct mechanistic link between epistatic coupling among residues and macroscopic evolutionary hysteresis.

An important implication for our findings on hysteresis in stochastic environments is that the sudden onset of environmental noise might disrupt the viability of populations with epistatic networks tuned for predictable environments. The dynamic adaptation of epistatic networks to account for greater environmental stochasticity might be difficult for some populations due to the underlying frustration of the physical molecules, such as interactions between amino acid residues in proteins that prevent proper folding \cite{starr_epistasis_2016,bryngelson_spin_1987}. Thus, the onset of high environmental variability due to climate change can lead to increased extinction for these populations \cite{duffy_climate-mediated_2022}, and populations with more moldable epistatic networks may fare better in these conditions \cite{kashtan_varying_2007,draghi_evolutionary_2009}, or perhaps simply populations with less epistatic constraints overall \cite{pinney_parallel_2021}.

Our findings suggest how epistasis, mutation rate, and environmental noise sensing might co‐evolve. Previous work on regulatory networks shows how cells can evolve architectures to buffer biochemical noise \cite{ozbudak_regulation_2002,thattai_intrinsic_2001}. Other work on fluctuating environments shows how phenotypic diversification or sensing strategies can evolve \cite{kussell_phenotypic_2005}. Meanwhile, theory of mutation rate evolution shows that mutator alleles can be selected in adapting populations \cite{tenaillon_second-order_2001}. Together, this suggests a plausible interplay: high‐mutation lineages may favor tighter noise‐buffering regulation, whereas lower mutation lineages may favor broader evolutionary sampling.

\section{Methods}
\subsection*{Wright-Fisher Model Simulation}
Here we describe the protocol used to simulate hysteresis loops for the epistatic Wright-Fisher model. Starting with a population with equal proportions of the four genotypes, we evolved the population for $1000$ generations with fixed values of $s$ and $J$. We varied $s$ in a quasi-static cycle every $1000$ generations from $0.5$ to $-0.5$ and back to $0.5$ to complete one evolutionary cycle. For Fig.\ref{fig:1}c, this scheme was performed independently for $N=(500,10000)$, $u=0.001$, and 20 values of $J$ linearly spaced from $0$ to $0.2$. We simulated $48$ independent replicate trajectories for each parameter set $(N,J,u)$, resulting in $18000$ total trajectories for analysis. For each trajectory, we computed the area between the upper ($s\rightarrow 0^+$) and lower ($s\rightarrow 0^-$) stationary points over a full selection cycle, which we call the hysteresis area. The stationary points were defined as the average frequency of the allele under selection in the last $250$ generations of a given leg of fixed selection strength $s$. The simulations of the two-locus Wright-Fisher model for linked loci with recombination were performed by a similar quasi-static selection cycling protocol (SI. II) with $N=10000$ and $u=0.01$, for $J=0,0.1,0.2$ and a range of recombination rates $r=0$-$0.5$. We reported the average hysteresis loop area from 48 replicate trials (Fig.S2)

The simulation results presented in Fig.\ref{fig:1}d were obtain as follows. We performed simulations of a one-dimensional Ornstein-Uhlenbeck process with advection and diffusion terms given by the terms $A(p)$ and $B(p)$, respectively (Eqn.S1), which are functions of the model parameters $s,J,u$ and $N$. Starting from the local fitness maximum $p_0$, the stochastic variable $p(t)$ was allowed to evolve until it reached the barrier crossing point given by $p_s$, and the time of first crossing was recorded. We performed 1000 independent trials of this process for a range of population sizes $N = 500$-$10000$ and selection strengths $s=0.2,0.225,0.25$, with $J=0.5$. We report the average rate of escape $\kappa_\textrm{esc} = 1/\langle T_\textrm{esc}\rangle$ with 95\% confidence intervals determined by propagation of errors.

The data presented in Fig.\ref{fig:2}a were obtained with simulation parameters $A = 0.5, \Omega = 0.001$, $\sigma = 0,0.1371$ and $J = 0,0.4$, and depict the average allele frequency of 48 independent replicate simulations. For the results presented in Fig.\ref{fig:2}b, we simulated 48 independent replicate trajectories of the epistatic Wright-Fisher model starting from equal genotype populations for model parameters $J=0$-$0.4$, $\sigma=0$-$0.1$, $u=0.0001$-$0.01$, $N=500$, $A = 0.5$, and $\Omega = 0.001$. This resulted in $19200$ total trajectories for analysis. The quantity $T$ in Eqn.\ref{fitness_shortfall} represents the total number of selection stages $s(t)$, which was chosen to complete one full hysteresis cycle for a given cycling rate $\Omega$. The results presented in Fig.\ref{fig:3} were obtained by 48 independent replicate trajectories for the following parameters: $J=0,0.03,0.1$, $N=500$, $u=0.01$, selection amplitude $A = 0.5$, cycling rate $\Omega = 0.0002$-$0.02$, and selection noise $\sigma = 0$-$0.1$.  The resulting allele frequencies around $s=0$ were fit to the effective fitness potential model by non-linear least squares optimization using the SciPy package with Python 3.9. Uncertainty in the fit parameters were estimated as the 95\% confidence interval of 500 non-parametric bootstrap simulations.

We performed simulations of the $L$-locus $K$-allele Wright-Fisher model (Eqn.\ref{KL}) with disordered epistatis to investigate the link between epistatic interaction networks and macroscopic hysteresis in long evolutionary cycles. The simulations were performed in a similar manner as for the two-locus two-allele investigations, involving slow cycling of selection pressure with fixed epistatic interaction parameters. First, we assigned single-site fitness parameters $s_i^\alpha$ to each allele type $i$ at site $\alpha$, with $s_i^\alpha$ ranging from 0 to maximum (or minimum) fitness value $s \in (-2,2)$. The epistatic parameters $J_{ij}^{\alpha,\beta}$ were generated randomly from a Gaussian distribution $\mathcal{N}(\langle J \rangle,\delta J)$ with mean $\langle J \rangle$ and variance $\delta J$. The simulations were run in sequential stages of fixed $s$ for 1000 generations each before moving on to the next selection target. At the end of each evolutionary stage, we calculated the response of the population by Eqn.\ref{fitness_response}.
\begin{equation}
    \bar{f}(\mathbf{p},s) = \frac{1}{L}\sum_l^L \sum_i^K s_i p_{il}
    \label{fitness_response}
\end{equation}
where $\bar{f}$ is the fitness response of the population evolving under selection $s$, $\mathbf{p} = {p_{11},p_{12},...,p_{21},...,p_{LK}}$, $s_i$ is the relative fitness contributed by allele type $i$, $p_{il}$ is the frequency of allele type $i$ among site $l$ of all individuals in the population, and $L$ is the total number of sites. This fitness response is plotted on the y-axis in Fig.\ref{fig:4}b (bottom row) and Fig.S5. When performing replicate simulations for the ``LK'' model, we regenerated the random Gaussian $J_{ij}^{\alpha,\beta}$ for each replicate, and the error bars in Fig.\ref{fig:4}b and Fig.S5 represent the standard deviation of 500 non-parametric bootstrap simulations based on the uncertainty from the replicate trials. The results presented in Fig.\ref{fig:4}b were performed with parameters $L=50$, $K=20$, $u=0.005$, $N=1000$, and $J=0.05$ for $\delta J = 0.015, 0.025, 0.05, 0.0625$. The parameters used for Fig.\ref{fig:6} are: Fig.\ref{fig:6}a: $L=20, K=20, u=0.01,N=1000, \langle J \rangle = 0,0.03,0.05,0.06$, $\delta J = 0$-$0.05$. Fig.\ref{fig:6}b: $L=5,20,50$, $K=20$,$u=0.002$,$N=1000$, $\langle J \rangle = 0.02$, $\delta J = 0$-$0.05$.

\subsection*{Protein Evolution Simulation}
To simulate the evolution of protein sequences under selective pressure and epistasis, we modified the Protein Evolution Parameter Calculation (PEPC, \cite{piszkin_extremophile_2021}) to include protein sequence flexibility prediction using the MEDUSA prediction software \cite{vander_meersche_medusa_2021}. This method classifies amino acid residues by contribution to local structural flexibility/rigidity. Since the MEDUSA tool was trained using homologous sequences and physico-chemical properties of amino acids, it necessarily incorporates epistatic interactions between residues. Fig.\ref{fig:4}a depicts the procedure for evolving protein sequences in the modified PEPC method. Every generation, a candidate amino acid substitution is determined from a BLAST query against the Pfam database and the BLOSUM80 matrix \cite{piszkin_extremophile_2021}. Then, a MEDUSA prediction is run on the candidate mutated sequence to assign a confidence level for the placement of each residue into one of five flexibility classes. To increase throughput, the HHblits MSA construction step of the MEDUSA prediction was performed only if the sequence had diverged by more than 80\% sequence identity from its sequence at the last MSA construction, otherwise the most recent MSA was used. The predicted global flexibility $F$ of the protein sequence was calculated from the MEDUSA results by Eqn.\ref{global_flex}, where $c_{l,a}$ is the confidence that residue $l$ belongs to the MEDUSA flexibility class $a$, and $n$ is the length of the protein.
\begin{equation}
    F = \sum_{l=1}^n(c_{l,1} + 2c_{l,2}+3c_{l,3}+4c_{l,4}+5c_{l,5})
    \label{global_flex}
\end{equation}
Here, the classes range from least flexibility class 1 to most flexible class 5. The global flexibility of the candidate sequence is compared to the parent sequence and the mutation is accepted with a probability determined by a Metropolis-Hasting criterion defined by Eqn.\ref{acceptance_prob}. 
\begin{equation}
    A(i\rightarrow j) = \min \left[ 1, \exp \left(-\frac{|F_i - F_j|}{T_{evo}}\right)\right]
    \label{acceptance_prob}
\end{equation}
In this scheme, a mutation is always accepted if it moves the global flexibility of the protein closer to a predefined target flexibility, while a deleterious mutations of $i \rightarrow j$ are accepted with probability $A(i \rightarrow j)$, where $F_i$ and $F_j$ are the global flexibilities of the parent and candidate sequences, respectively, and $T_{evo}$ is a mutation "temperature" factor that modifies the purity of the selective pressure. When $T_{evo}=0$, only beneficial mutations are accepted, while increasing $T_{evo}$ from 0 increases the likelihood that deleterious mutations are accepted. After a mutation is accepted or discarded, a new candidate mutation is chosen for the new sequence and the process is repeated for the full evolutionary cycle.

\quad We performed the PEPC simulations for the wild-type sequences of Cold Shock Protein A in \textit{Escherichia coli}, Profilin-2 in \textit{Olea europaea}, TEM $\beta$-lactamase from \textit{E. coli}, and $\beta$-galactosidase in \textit{E. coli} (UniProt Ascension IDs:  P0A9X9, A4GDR8, P62593 and P00722, respectively). For each simulation, we defined a list of target global flexibilities and evolved the sequence in successive stages for each target. Each stage of evolution was run until the target flexibility was reached or 500 generations had passed. At the end of each stage, we collected the final global flexibility of the sequence, then moved the target flexibility and began the next stage. This was repeated in a cycle until the target flexibility returned to its initial value. For each protein we performed 20 independent trials of full evolutionary cycles with $T_{evo} = 0$ and calculated the aggregate statistics to visualize the trajectories depicted in Fig.\ref{fig:4}b, which represent the mean (points) and standard deviation (error bars) of the replicates. Optionally, AlphaFold structural predictions can be performed on sampled sequences of the full trajectory to check if the simulation deviates from a reasonable folded structure. For our simulations, the structures predicted by AlphaFold 2 maintained average predicted local distance difference test (pLDDT) scores above 70 for sample sequences across the entire trajectories, with the exception of CspA which dipped slightly below 70 (Fig.S4). A pLDDT score above 70 indicates high confidence in the predicted backbone fold \cite{jumper_highly_2021}. 

\section{Author Contributions}
DCV and LP conceptualized the idea. DCV and LP derived analytical results. LP conducted numerical simulations. LP and DCV prepared figures and wrote the manuscript.

\section{Materials \& Correspondence}
Correspondence and requests should be addressed to DCV.  

\section{Data Availability}
Original data generated in this study is accessible from the Zenodo repository: 10.5281/zenodo.18001835. Public data used in this study are referenced in the article.   

\section{Code Availability}
The code developed for simulation and analysis is available in the Zenodo repository: 10.5281/zenodo.18001835. Additional simulation and analysis scripts are also available from the following Github repositories: LukePiszkin/Wright-Fisher-Simulation, /Protein-Evolution-Hysteresis, and /Lake\_Mendota\_hysteresis. All original code generated for results presented in the study are included in the article, supplementary information, or Zenodo repository. 

\section{Competing Interests Statement}
The authors have no competing interests to declare.

\section{Acknowledgments}
LP would like to thank Alan Lindsay and Kobby Van Dyck for stimulating discussions. This research was supported in part by the Notre Dame Center for Research Computing. We specifically acknowledge the assistance of Dodi Heryadi and Scott Hampton for technical support.

\bibliographystyle{unsrt}
\bibliography{references}

\end{document}